\documentclass{article}
\usepackage{graphicx} 
\usepackage{booktabs}
\usepackage{hyperref}

\usepackage[a4paper, margin=1.4in]{geometry}  

\def\eg{\emph{e.g.~}}
\def\etal{{\em et al.~}}

\title{How good is the h-index?}
\author{Ali Borji \\
aliborji@gmail.com}

\date{January 2025}

\begin{document}

\maketitle

\begin{abstract}
    The h-index has become a widely used metric for evaluating the productivity and citation impact of researchers. Introduced by physicist Jorge E. Hirsch in 2005, the h-index measures both the quantity (number of publications) and quality (citations) of a researcher's output. While it has gained popularity for its simplicity and practicality, the h-index is not without its limitations. We examine the strengths and weaknesses of this metric, presenting preliminary experimental results that demonstrate the limitations of the h-index. We also propose a potential solution. The primary aim of this work is to shed light on the shortcomings of the h-index and its implications for ranking scientists, motivating them, allocating funding, and advancing science.
    
\end{abstract}

\section{Introduction}

The h-index~\cite{Hirsch2005} serves as a representation of a researcher's impact throughout his research journey, playing a crucial role in advancing his career and contributing to the comparative assessment of academic institutions. Considering the widespread use and significant influence of this index in academia, it prompts a moment of reflection: Is this truly the most effective approach for evaluating the merit of individual scientists? Are there better alternatives?

The h-index is considered to be one of the most important features of a researcher's reputation. In fact, entire careers can be boiled down to this one number. The h-index is important because it directly influences access to funding and resources. Researchers with higher h-indices are often viewed as more impactful, increasing their likelihood of securing support from funding agencies.

The h-index offers several advantages. It combines productivity and impact, balancing publication quantity and citation quality. It is simple and easy to interpret. It reduces the influence of outliers by focusing on consistent impact rather than single highly-cited papers. It allows for cautious comparisons within disciplines, and it reflects career progression over time.

The h-index, as a quantitative measure of academic output, is susceptible to manipulation. Coercive citation practices, where journal editors demand authors include unnecessary citations to their own work, can artificially inflate the h-index. Self-citations and even computer-generated documents can also be employed to manipulate the index, particularly when using platforms like Google Scholar. Moreover, hyper-authorship\footnote{Hyper-authorship refers to the phenomenon of scientific papers having an unusually large number of co-authors, often in the dozens or even hundreds.} can skew the h-index by assigning disproportionate credit to individuals on papers with a large number of authors. Recent studies indicate a decline in the correlation between the h-index and awards that traditionally signify recognition by the scientific community~\cite{koltun2021h}.

Several studies have discussed the strengths and limitations of the h-index in evaluating academic performance \cite{bornmann2009state,bornmann2007we,waltman2012inconsistency,jacso2008pros}. Here, we contribute by presenting results of a vote.

\begin{figure}
    \centering
    \includegraphics[width=0.45\linewidth]{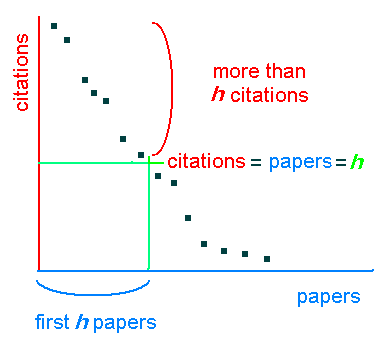}
    \includegraphics[width=0.45\linewidth]{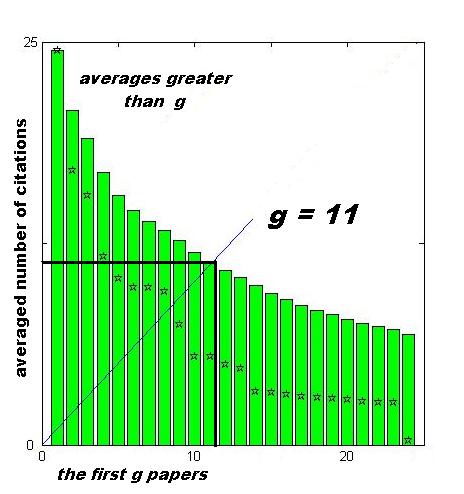}
    \caption{Left: \href{https://en.wikipedia.org/wiki/h-index}{h-index}, Right: 
    \href{https://en.wikipedia.org/wiki/g-index\#cite\_note-Egghe-1}{g-index}. See text for details.}
    \label{fig:indices}
\end{figure}

\section{What is the h-index?}

The h-index is calculated by counting the number of publications for which an author has been cited by other authors at least that same number of times. For instance, an h-index of 17 means that the scientist has published at least 17 papers that have each been cited at least 17 times. Please see Fig.~\ref{fig:indices}.

The h-index is a metric that measures an author's productivity and citation impact, initially for individuals but now also applied to journals and groups of scientists. Proposed by physicist Jorge E. Hirsch in 2005~\cite{Hirsch2005}, it correlates with success indicators like Nobel Prizes and research positions.

However, some have criticized the h-index concept for not exactly being the best metric to analyze someone's research importance. For example, one Quora user says that h-index is highly dependent on the field, and that it is perhaps not the best benchmark.

Metrics akin to the h-index have found application beyond the assessment of authors or journals. For instance, the h-index has been extended to evaluate YouTube channels. In this context, it represents the number of videos with at least h multiplied by $10^5$ views. Comparing this index with the total view count of a video creator provides a more comprehensive assessment of both productivity and impact within a single metric. Moreover, a derivative Hirsch-type index has been developed for institutions. This index measures the success of a scientific institution based on the number of researchers from that institution with an h-index equal to or greater than the index value.

\subsection{Alternatives to the h-index}
\noindent \textbf{The pagerank-index.}
Senanayake \etal~\cite{Senanayake2015} propose using the page-rank algorithm to rank publications in the citation network. Each node gets a value after this process, which can then be distributed to each author, and the summation of all page-rank values is obtained for every author. This can then be compared to all other author values to form the percentile.
The advantage of doing so, is that PageRank can compare the sources of information and determine which references are more-trustworthy. PageRank is calculated recursively and depends on the metric of all pages that link to it. Each page spreads it vote equally among all out-links. If a page is linked to by many high ranked pages, it achieves a high rank. Here, not all citations are equal, and a publication is important if it is pointed to by other important publications.

\noindent \textbf{The g-index.}
The g-index is an author-level metric suggested in 2006 by Leo Egghe (Fig.~\ref{fig:indices}; right panel). The index is calculated based on the distribution of citations received by a given researcher's publications, such that given a set of articles ranked in decreasing order of the number of citations that they received, the g-index is the unique largest number such that the top g articles received together at least $g^2$ citations. Hence, a g-index of 10 indicates that the top 10 publications of an author have been cited at least 100 times ($10^2$):

\begin{equation}
    g^2 \leq \sum_{i \leq g} c_i
\end{equation}

It can be equivalently defined as the largest number n of highly cited articles for which the average number of citations is at least n.

\section{How good is the h-index?}

This section scrutinizes limitations of h-index, such as its susceptibility to manipulation through self-citations and its bias towards established researchers over early-career academics.

\subsection{Shortcomings of the h-index}
The h-index fails to account for certain nuances in evaluating researchers' contributions. Some of the following failures are not exclusive to the h-index but rather shared with other author-level metrics:

\begin{enumerate}
    \item It exhibits implicit age bias, as it may rate a young scientist with a few highly-cited papers equivalently to a more experienced one with numerous less impactful publications, undermining meritocracy.

    \item It overlooks the prestige of the journals in which scientists publish, treating all citations equally regardless of source authority. This disregards the significance of citations from top-tier journals versus lesser-known ones.
    
    \item It incentivizes quantity over quality, potentially leading researchers to prioritize producing a greater number of publications to inflate their index rather than focusing on the depth and significance of their scientific contributions. It ranks a researcher with one or two groundbreaking works cited thousands of times lower than someone with many moderately cited papers, which may lead to undervaluing truly transformative contributions. We will discuss this in more detail in the next subsection.

    \item It does not distinguish between lead authorship and minor contributions. A researcher who contributes marginally to many high-impact papers may have a higher h-index than someone with groundbreaking solo-authored work.

    \item It does not account for the number of authors of a paper. In the original paper, Hirsch suggested partitioning citations among co-authors. One such fractional index is known as h-frac, which accounts for multiple authors but is not widely available through the use of automatic tools.
    
    \item It discards the information contained in author placement in the authors' list, which in some scientific fields is significant though in others it is not.

    \item It is an integer, which reduces its discriminatory power. Ruane and Tol~\cite{ruane2008rational} therefore propose a rational h-index that interpolates between h and h + 1.
    
    \item It does not account for the different typical number of citations in different fields, \eg experimental over theoretical. Citation behavior in general is affected by field-dependent factors, which may invalidate comparisons not only across disciplines but even within different fields of research of one discipline.

    \item It is not universally applicable across fields. Mathematicians and physicians typically receive fewer citations on average.

    \item It is confined to published and cited work, overlooking other contributions to science, such as teaching, mentorship, public engagement, patents, or software development.
    
    \item It varies depending on the database (\eg Google Scholar, Scopus, or Web of Science) due to differences in citation coverage. This inconsistency can make comparisons unreliable.

    \item It can be inflated by researchers through self-citations or by participating in citation networks where colleagues excessively cite each other's work.

\end{enumerate}

\begin{figure}[t]
    \centering
    \includegraphics[width=0.5\linewidth]{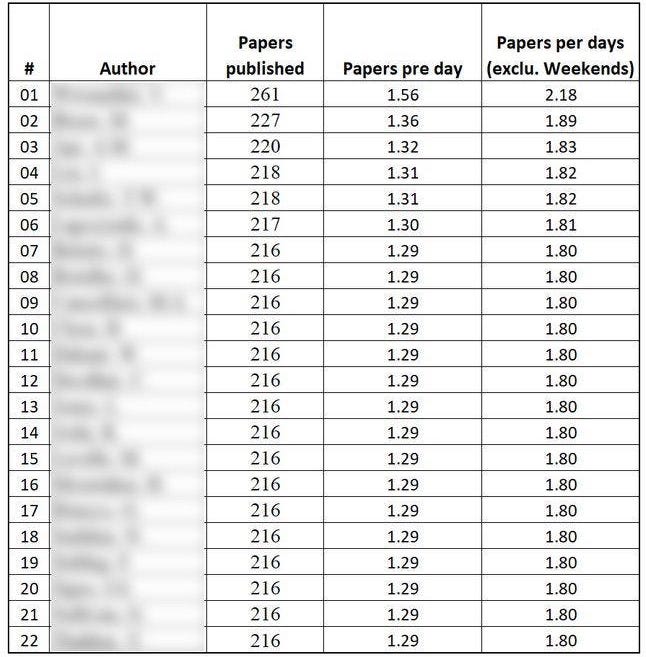}
    \caption{In the first half of 2024, according to Scopus, 22 authors have each published over 200 papers! \href{https://x.com/fake_journals/status/1802151185480307132}{Source}.}
    \label{fig:pubs}
\end{figure}

\subsection{The h-index encourages quantity over quality}
Assigning credit is generally complex, and it becomes even more challenging in the context of scientific contributions. Ideally, an author of a scientific paper should have invested a significant amount of intellectual effort into the publication. However, this is often not the case. In the modern era, it is not uncommon for some authors listed on a paper to have contributed nothing to the publication. To illustrate this, consider the following statistic:
In the first half of 2024, according to Scopus, 22 authors have each published over 200 papers. Fig.~\ref{fig:pubs} provides the details (author names are not shown). The most prolific author has published 261 papers as of June 15, 2024, averaging 1.56 papers per day. Excluding weekends, this translates to 2.18 papers per day!

The issue with current indices for measuring academic contributions is that they do not penalize this behavior. Admittedly, this concern extends beyond just indices, but it remains a significant problem.

\subsection{The h-index of famous scientists}

While the h-index is a useful metric for measuring productivity, it does not fully capture the impact or genius of a scientist. Many of the most influential scientists in history are known for one or two groundbreaking works that changed the world, even if their h-index does not reflect it.

Table~\ref{tab:hindices} shows how the h-index applies to some famous scientists, along with examples of their groundbreaking work. To provide context and facilitate comparison, consider the h-indices of prominent AI researchers such as Yoshua Bengio (241), Geoffrey hinton (188), and Yann LeCun (154). Interestingly, many lesser-known researchers have h-indices exceeding those of the scientists listed in Table~\ref{tab:hindices}. It is worth noting that the data in Table~\ref{tab:hindices} may not be entirely accurate and is presented solely for illustrative purposes.

Unlike past scientists who often conducted a limited number of groundbreaking works independently, contemporary researchers benefit from collaborative efforts, involving students and large teams, to significantly boost their h-indices.

\begin{table}[t]
\centering
\footnotesize
\begin{tabular}{lcl}
\toprule
\textbf{Scientist}       & \textbf{h-index} & \textbf{Famous Work(s)} \\
\midrule
Albert Einstein          & ~157              & Special/General Relativity, Photoelectric Effect \\
Richard Feynman          & ~62              & Quantum Electrodynamics, Feynman Diagrams \\
John Nash                & ~13              & Nash Equilibrium (Game Theory) \\
Stephen Hawking          & ~135              & Hawking Radiation, Theoretical Cosmology \\
Francis Crick            & ~92              & Discovery of the DNA Double Helix \\
Alan Turing              & ~45              & Turing Machine, Cryptanalysis, Turing Test \\
John von Neumann         & 102                 & Game Theory, Von Neumann Architecture   \\
& &  Quantum Mechanics \\
\bottomrule
\end{tabular}
\caption{The h-indices and major contributions of some famous scientists. Data from Google scholar.}
\label{tab:hindices}
\end{table}

\section{Our study}

\begin{figure}
    \centering
    \fbox{\includegraphics[width=.8\linewidth]{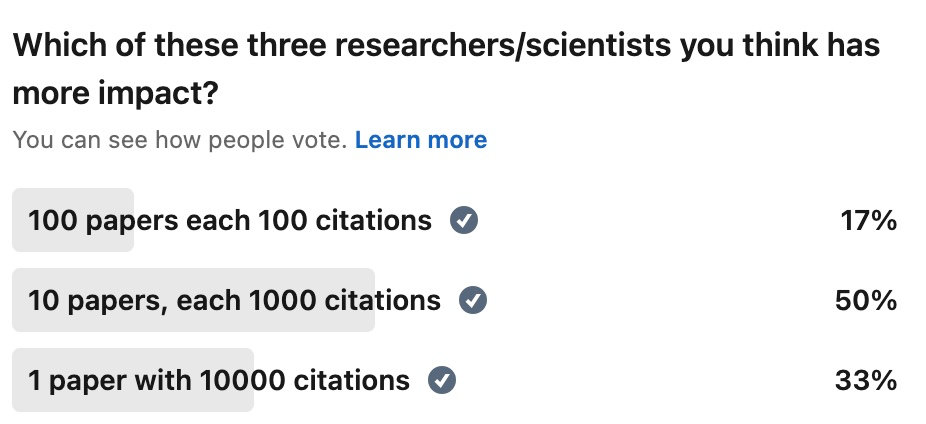}}
    \caption{Poll results from 83 voters in LinkedIn. The h-index for these authors in order are 100, 10, and 1.}
    \label{fig:exp}
\end{figure}

We undertook an informal investigation on LinkedIn, inviting individuals to respond to the inquiry shown in Fig.~\ref{fig:exp}. The bulk of the voters (50\%) chose a compromise, suggesting that there are few or just enough works with a high number of citations. A researcher with numerous papers (option I), each with few citations, was not preferred, suggesting a lack of expertise in specific areas. The third option was not favored either, suggesting the author might have had one successful paper but has not consistently produced high-impact work. However, the ultimate winner was the second option, striking a balance between the first and third choices by presenting a sufficient number of highly cited works.

\section{Discussion and conclusion}
Assessing the contributions and influences of individuals across various fields such as science, innovation, engineering, business, politics, literature, music, art, and more is a complex task. While some individuals achieve recognition for a single notable achievement or creation, others may be renowned for their consistent production of smaller works. Therefore, there is not a one-size-fits-all approach to evaluating success. The h-index, while offering a simple numerical metric, may not be sufficient due to the aforementioned reasons.

The h-index is undoubtedly a valuable tool for assessing academic impact, but it should not be used in isolation. A researcher’s contributions to science and society are multifaceted, encompassing more than just publications and citations. Supplementary metrics, such as the I10-index (papers cited at least 10 times; similarly, the I1000-index), total citations, or field-normalized impact scores, can provide a more holistic view. Moreover, qualitative assessments, such as peer reviews, awards, and mentorship roles, are crucial for capturing a researcher’s broader impact. Over-reliance on metrics like the h-index risks oversimplifying the diverse ways in which researchers contribute to their fields.

\subsection{A possible solution}
We suggest exploring the idea of training machine learning algorithms to emulate the evaluations made by researchers or through a voting mechanism. A relevant example is the use of perceptual loss in developing loss functions and similarity measures for images, which has been trained on human judgments~\cite{Johnson2016}. Similarly, we could consider applying a comparable approach here: using human rankings of authors based on their paper citations to train the model. This approach can be particularly valuable in cases where human expertise or judgment is difficult to replicate algorithmically.

\subsection{What defines a true scientist?}
We believe a true scientist is one who is deeply committed to the pursuit of knowledge, making direct contributions through personal effort, intellectual rigor, and innovation. Historical figures like Isaac Newton and Albert Einstein exemplify this ideal, as their groundbreaking discoveries were the result of intense reflection, experimentation, and a relentless drive for understanding. These individuals embodied science as a deeply personal endeavor, driven by curiosity and a commitment to uncovering truth through their own intellectual labor. You are invited to read more on this topic at: \href{https://medium.com/@aliborji/what-defines-a-true-scientist-reflections-on-hintons-nobel-prize-win-46ff9c280514}{Here}.

In contrast, modern science operates within a collaborative and systematized framework, where some researchers focus more on administrative tasks, funding, and advising rather than hands-on discovery. While this approach has accelerated progress and increased productivity, it has also blurred the distinction between those who lead and those who contribute directly to the science. A true scientist, however, is one whose work reflects personal engagement, deep understanding, and a genuine curiosity to solve problems. Despite the pressures of modern academia, the essence of being a scientist remains rooted in the intrinsic excitement of discovery, echoing the spirit of the great thinkers of the past. An effective index for evaluating researchers' contributions and rankings should incorporate these factors and actively discourage practices aimed at artificially inflating the metrics.

\bibliographystyle{plain}
\bibliography{refs}

\end{document}